\documentclass[11pt,a4paper]{article}
\pdfoutput=1
\usepackage{jheppub}
\usepackage{graphicx,amssymb,amsmath,amsfonts}
\usepackage[english]{babel}

\newcommand{\ep}{{\epsilon}}

\makeatother

\title{On the dual flow of slow-roll Inflation}
\author{Uri Kol}

\affiliation{School of Physics and Astronomy,\\
The Raymond and Beverly Sackler Faculty of Exact Sciences,\\
Tel Aviv University, Ramat Aviv 69978, Israel}

\emailAdd{urikol@post.tau.ac.il}

\abstract{
We study the dual 3d Euclidean RG flow of single-field slow-roll Inflation using the postulates of the dS/CFT correspondence. For that purpose we solve for the inflationary fluctuation at all times using a matching procedure between two approximate solutions which are separately valid at different regions of the space of parameters but together cover all of it. The two modes of the full solution mix such that each of the modes at late times is a superposition of the modes in the quasi-de Sitter region. We find that the dual theory admits two phases of explicit and spontaneous breaking of conformal symmetry.

We also find that the mixing effect between the two modes in the bulk implies that slow-roll inflation does not guarantee, but rather generically generates, a nearly scale invariant power spectrum, except in fine-tuned situations. We suggest that the mixing effect can have a unique signature on other cosmological observables such as the bispectrum.
}

\begin{document}
\maketitle

\section{Introduction and summary of the results}

The $dS/CFT$ correspondence was first proposed in \cite{Witten:2001kn,Strominger:2001pn} and further developed in \cite{Maldacena:2002vr} for single field inflationary models. In analogy with the AdS/CFT correspondence, it suggests a duality between a four dimensional gravity theory which is asymptotically de Sitter and a three dimensional Euclidean conformal field theory. More concretely it suggests the following dictionary
\begin{eqnarray}\label{dictionary}
  \Psi[g] &=& Z[g]
\end{eqnarray}
$\Psi[g]$ is the wavefunction of the universe for a given three metric $g$ and $Z[g]$ is the partition function of the dual conformal field theory.

According to the inflationary paradigm the universe starts its life in a de Sitter phase and then evolve, due to various matter fields, until it reaches a second de Sitter phase at asymptotically late times.
The primordial and final de Sitter phases are characterized by two different Hubble constants $H$ and $H_0$, respectively.
The isometries of de Sitter space correspond to the conformal symmetry group of the dual field theory and therefore the evolution of the universe can be described, using the dS/CFT correspondence, as a renormalization group (RG) flow between two conformal fixed points with central charges \cite{Strominger:2001gp,
Larsen:2002et,Larsen:2003pf,vanderSchaar:2003sz,Larsen:2004kf,Schalm:2012pi,Bzowski:2012ih,McFadden:2013ria,Kiritsis:2013gia}.
\begin{eqnarray}
  c_{IR} &=& H^{-2} \\
  c_{UV} &=& H_0^{-2}
\end{eqnarray}
respectively\footnote{For a different relation between inflation, field theory and holography see \cite{Nastase:2011qz}, where a cosmological brane inflation scenario was constructed using a holographic MQCD model.}.
The primordial de Sitter corresponds to the IR fixed point and the final de Sitter correspond to the UV fixed point.
This is a manifestation of the dS/CFT correspondence as a UV/IR duality.
Note that since the final Hubble constant $H_0$ is much smaller than the primordial one $H$, the central charges exhibit the inequality $c_{UV}> c_{IR}$, as appropriate for a consistent RG flow.
Moreover, a monotonically decreasing function along the flow (or monotonically increasing with time) can be constructed and therefore a $c$-theorem for cosmology can be established \cite{Sinha:2010pm}.

The dual field theory to gravity in anti-de Sitter space is known to be $\mathcal{N}=4$ $SYM$ and the AdS/CFT correspondence is well-established \cite{Maldacena:1997re,Aharony:1999ti} (including symmetry breaking processes and geometries dual to RG flows, see for example \cite{Klebanov:1999tb,Freedman:1999gp}).
However, much less is known about the dS/CFT correspondence and in particular the dual field theory to gravity in de Sitter space, if exists, is not known.
Therefore it is interesting to study and characterize the dual theory to gravity in de Sitter space using the proposed dictionary \eqref{dictionary}.
But it is also possible to use the dictionary in the opposite direction, to study gravity in de Sitter using the dual theory (or more precisely, its symmetries). For example, the conformal symmetry has been used to constrain the form of the three-point function \cite{Maldacena:2011nz,Mata:2012bx} and the Maldacena consistency consitions \cite{Maldacena:2002vr} were rederived using the Ward identities of the dual field theory \cite{Schalm:2012pi}.

In this paper we study single field slow-roll inflation, in which the universe departs from its primordial de Sitter phase due to the potential of a single scalar field, the inflaton.
Since the primordial de Sitter phase of the universe is dual to the IR fixed point, the dual process to the departure from this phase is described by a perturbation of the dual conformal theory by a marginally irrelevant operator \cite{Larsen:2002et,Larsen:2003pf,vanderSchaar:2003sz}. The spectrum of perturbations caused by the inflaton is therefore dual to the spectrum of perturbations caused by the marginally irrelevant operator.
The aim of this paper is to study the dual spectrum of perturbations in a similar manner to the study of holographic RG flows in AdS \cite{Hoyos:2012xc,Hoyos:2013gma}.
In order to use the dictionary \eqref{dictionary} in the same way as in AdS/CFT one has to calculate the fluctuation in the bulk at all times. On the one hand, the dual field theory lives on a spatial slice at asymptotically late times (analogous to the AdS boundary), and on the other hand, the Bunch-Davies boundary conditions have to be imposed on the solution at early times.
However, the solution for the fluctuation at all times will not only be useful for the study of the dual theory but also to study the bulk gravity, namely inflation.
In general, it is not possible to find the solution for the fluctuation at all times, but we will be able to do this in a low momentum approximation using a technique that was developed to understand holographic RG flows between two fixed points.
The technique is called the \emph{matching procedure} and we will describe it below.

\begin{figure}
  \center \includegraphics[scale=0.6]{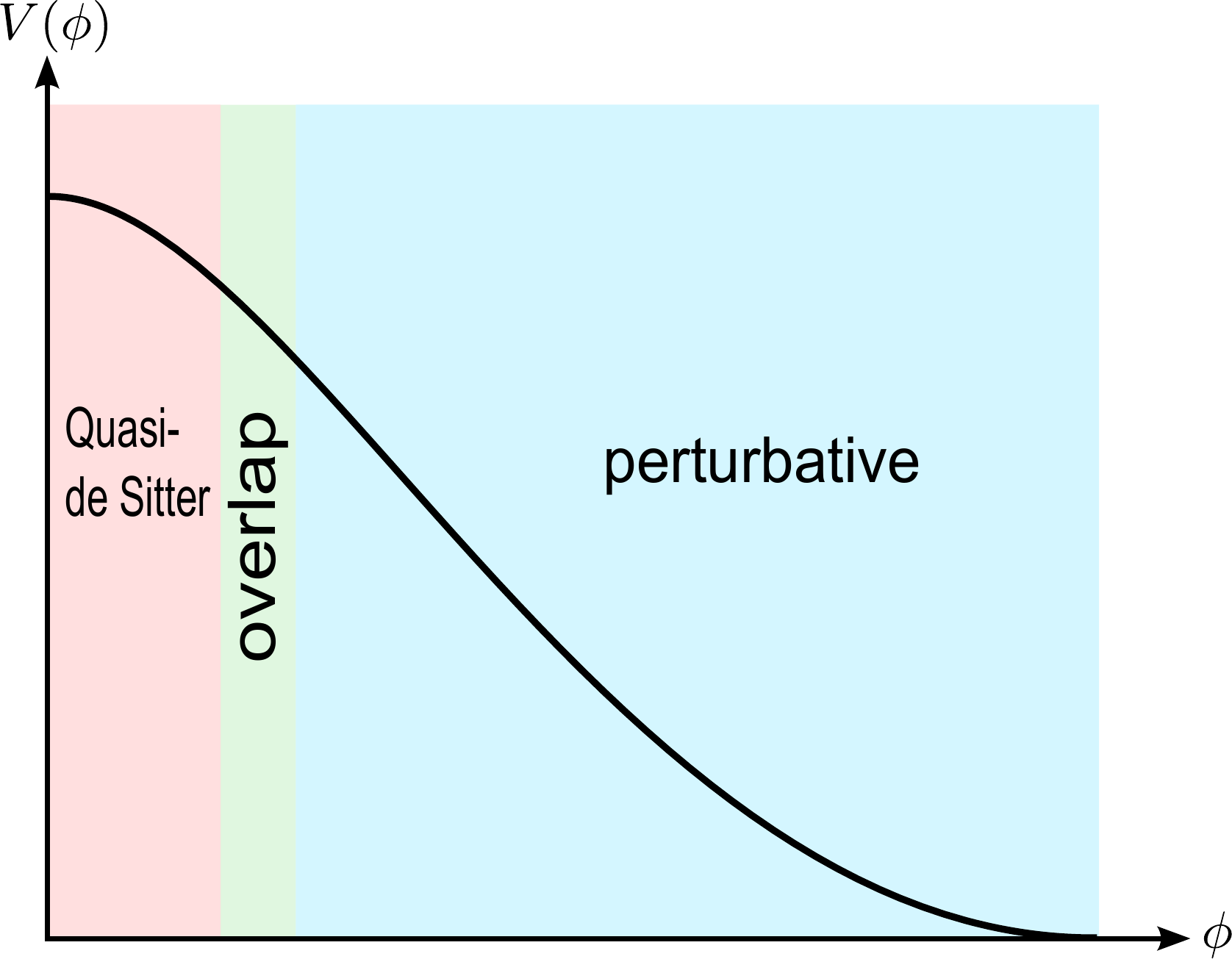}\\
  \caption{The inflationary potential with two critical points. The maximum on the left represents the primordial de Sitter space while the minimum on the right is the asymptotically late de Sitter. The region around the maximum is called quasi-de Sitter. In the quasi-de Sitter region one can solve the equation of motion of the fluctuation in a slow-roll expansion.
  On the other hand, one can solve the equation of motion perturbatively in small momenta. This perturbative solution is not valid at early times when the comoving Hubble radius is large compared to the wavelength of the fluctuation. However, for small enough momenta the perturbative solution will be valid all the way to the quasi-de Sitter region such that the two solutions overlap.
	}
	\label{potential}
\end{figure}

The superhorizon region is defined to be where the comoving Hubble radius of the universe, $\mathcal{R}$, is smaller than the wavelength of the fluctuation
\begin{equation}
  \frac{ \mathcal{R} } {\lambda} <1
\end{equation}
It is therefore possible to solve the equation of motion for the fluctuation, in the superhorizon region, perturbatively in large wavelength (or small momentum).
Since the comoving Hubble radius decreases during inflation, the approximation becomes more accurate at late times but loses its validity in the sub-horizon region at early times.
On the other hand, at early times the universe is a quasi-de Sitter space and one can solve for the fluctuation using the standard slow-roll prescription, with the Bunch-Davies boundary conditions.
While the perturbative solution will always break down at early enough times, for sufficiently large wavelength (or small momentum) its region of validity can extend all the way to the quasi-de Sitter region. Hence, for fluctuations of sufficiently large wavelength there is an overlapping region where both the low-momentum and the slow-roll approximations are valid. We are then able to construct a solution which is valid at all times by matching the two solutions in the overlapping region. This procedure is described in figure \ref{potential}, where we schematically describe the inflaton's slow-roll potential that interpolates between the two fixed points (corresponding to the two de Sitter spaces).

We find that the dual field theory admits two phases, depending on the solution for the homogenous background field. Since the inflaton field admits a second order differential equation, it has two solutions.
The solution which goes faster to zero at late times corresponds to a phase where conformal symmetry is spontaneously broken, while the other solution corresponds to an explicitly broken phase. This is of course natural in the language of AdS/CFT \cite{Hoyos:2012xc,Hoyos:2013gma}.
The spectrum of perturbations is nearly scale invariant in the explicit breaking phase. That spectrum corresponds to a perturbation by a nearly marginal operator. The anomalous dimension of that operator is related to the slow-roll parameters.
In the spontaneous breaking phase, on the other hand, we observe a massless pole which we associate to the propagation of the dilaton, the Goldstone boson of spontaneously broken conformal symmetry.

From the bulk point of view the phase structure of the dual field theory is related to a question of fine-tuning.
In general, both solutions for the homogenous background inflaton exist.
Hence, the generic case corresponds to the explicit breaking phase and therefore will generate a nearly scale invariant (or quasi-marginal) spectrum of perturbations.
However, when the leading solution for the homogenous background field at late times is fine-tuned to zero, the dual field theory will admit a spontaneous breaking phase and the power spectrum of perturbations, as a function of momentum, will behave as
\begin{equation}
  \mathcal{P} \sim k^2
\end{equation}
That behavior of the power spectrum corresponds to the appearance of a massless dilaton pole in the two-point function of the dual field theory.
The fact that we observe a nearly scale invariant spectrum of perturbations means that there is no fine-tuning in the universe, in the sense discussed above.

The appearance of the massless dilaton pole in the spontaneous breaking phase is directly related to a phenomenon of a mixing between the two fluctuating modes in the bulk and which was recently discovered in AdS/CFT \cite{Hoyos:2013gma,Bajc:2013wha}.
The matching procedure discussed above mix between the normalizable and the non-normalizable modes of the fluctuation in the bulk.
The meaning is that \emph{each of these modes at late times (end of inflation) is a superposition of the two modes in the quasi-de Sitter region}.

It was shown in \cite{Hoyos:2013gma} that generically (namely for general boundary conditions and shapes of the potential) there is a mixing between the two modes in the bulk. The values of the mixing coefficients depend on the precise details of the problem, and it is possible to tune some of them to zero, but generically they appear.
In this paper we choose the standard Bunch-Davies boundary conditions and a slow-roll potential, but the mixing effect does not depend on this choices.

The mixing effect between the two modes in the bulk may potentially affect cosmological observables.
While the non-normalizable mode stays constant in the superhorizon region, the normalizable mode decays exponentially fast. By the time the fluctuation re-enters the horizon after inflation ends, the normalizable mode have practically decayed to zero. The amplitude of the non-normalizable mode at horizon re-entry is then projected to the cosmological observables at later times using the transfer function.
Without the mixing effect, the amplitude of the non-normalizable mode at horizon exit and horizon re-entry is the same, and the perturbations will be described by the well-known nearly scale invariant power spectrum of single-field inflation.
However, if the two modes mix in the superhorizon region, as our solution shows, the amplitude of the non-normalizable mode at horizon re-entry will be different than its value at horizon exit, and that will affect the observed spectrum.
We find that in the explicit breaking phase, to leading order in momentum, the spectrum is nearly scale invariant and the mixing affects only at subleading orders.
In the spontaneous breaking phase, on the other hand, we find that the mixing effect generates a new leading term in the spectrum which is not nearly scale invariant, and hence dramatically changes the result. This term corresponds to the massless dilaton pole in the dual two-point function.
These results are in agreement with the measured nearly scale invariant spectrum since the spontaneous breaking phase corresponds to a fine-tuned situation, while the generic, non-fine-tuned, case corresponds to the explicit breaking phase, as discussed above.

Since the results that we present in this paper are in a complete agreement with the measurements, they are not of a direct phenomenological interest.
However, they suggest interesting possible effects.
We have discussed the power spectrum of the perturbations but it is not known how the mixing affects other cosmological observables such as higher order correlators.
It may very well be that, as opposed to the power spectrum, other cosmological observables are significantly affected also in the generic, explicit breaking, case.
Therefore, it would be very interesting to study the effect of the mixing on the bispectrum (or other cosmological observables) and to see, for example, whether it can cause a suppression of non-gaussianities.
This question is out of the scope of this paper and we leave it for further investigation.


The paper is organized as follows. In section 2 we briefly review single field inflation and the homogenous background solution. In section 3 we study scalar fluctuations over the homogenous background and construct the low momentum solution at all times using the matching procedure. Section 4 is devoted for the dual flow. We compute the two-point function for the dual operator to the inflaton field and discuss different phases of the dual field theory. In section 5 we study gravitational waves (tensor fluctuations). We conclude in section 6 and propose future directions.

\section{Single field inflation}

We study single field inflationary models described by the Lagrangian
\begin{eqnarray}
  S &=&  \int dt d^dx \sqrt{-g} \left[ \frac{1}{2}R -\frac{1}{2} \partial_{\mu}\phi\partial^{\mu}\phi -V(\phi) \right]
\end{eqnarray}
We have set $M_{\emph{pl}}^{-2}\equiv8 \pi G_N =1$. The equations of motion for the metric and the scalar are
\begin{eqnarray}
  \nonumber R_{MN} &=& \partial_{M}\phi\partial_{N}\phi +\frac{2}{d-1}g_{MN}V(\phi) \\
  \square \phi &=& \frac{\partial V}{\partial \phi}
\end{eqnarray}
The homogeneous solution to these equations has the form
\begin{eqnarray}\label{homogenousSol}
  \nonumber ds^2 &=& -dt^2 + e^{2A(t)}d\vec{x}^2 \\
  \phi &=& \phi(t)
\end{eqnarray}
where $\phi(t)$ and $A(t)$ obey the equations
\begin{eqnarray}\label{eom}
  \nonumber \ddot{A} &=& -\frac{1}{d-1} \dot{\phi}^2 \\
  \dot{A}^2 &=& \frac{1}{d(d-1)}\dot{\phi}^2 +\frac{2}{d(d-1)}V \\
  \nonumber \ddot{\phi} +d \dot{A} \dot{\phi} &=& -\frac{\partial V}{\partial \phi}
\end{eqnarray}
The third equation follows from the first two.

We are interested in potentials interpolating between two fixed points, as described in figure \ref{potential}. In general, it is not possible to find an exact solution to \eqref{eom} without specifying the exact form of the potential. However, it is possible to study the asymptotic forms of the solution near the fixed points. Let us start by looking exactly at a fixed point of the potential, where its derivative is zero. At this point the potential takes a constant value
\begin{equation}\label{constantV}
    V=\frac{d(d-1)}{2}H^2
\end{equation}
and the solution for \eqref{eom} at this point is
\begin{eqnarray}
  \phi(t) &=& \phi _0 \\
  A(t) &=& H t
\end{eqnarray}
where $\phi_0$ is some constant. This solution is a \emph{de Sitter} space with a Hubble constant $H$. The value of the potential at each of the fixed points is different and therefore the potential interpolates between two \emph{de Sitter} spaces with different Hubble constants.
Next, we want to study the behavior of the solution in the vicinity of the fixed point. Near a fixed point the potential can be expanded as follows
\begin{equation}\label{PotEx}
  V = \frac{d(d-1)}{2} H^2 + \frac{1}{2} m^2 (\phi-\phi_0) ^2
\end{equation}
Plugging this into \eqref{eom}, and defining $\delta \phi \equiv \phi -\phi_0$, we find the following equation for the scalar
\begin{equation}\label{scalarEOM}
  \ddot{\delta \phi} + d H \dot{\delta \phi} +m^2 \delta \phi =0
\end{equation}
The two solutions to \eqref{scalarEOM} are
\begin{equation}\label{scalarSol}
  \delta \phi _{\pm} = C e^{-\Delta_{\pm}Ht}
\end{equation}
where $\Delta_{+}$ $\left(\Delta_{-}\right)$ is the larger (smaller) root of the equation
\begin{equation}\label{roots}
  \Delta \left( \Delta -d  \right) = - \left( \frac{m}{H} \right)^2
\end{equation}
and $C$ is an arbitrary constant.
The scale factor is then
\begin{equation}\label{scaleFactor}
  A_{\pm} = Ht - \frac{C^2}{4(d-1)}e^{-2\Delta_{\pm} H t}
\end{equation}
Note that the value of $m^2$ is different in each fixed point (and, in particular, is negative for a maximum of the potential).

\section{Fluctuations}

In this section we study fluctuations over the homogenous background solution discussed in the previous section. We will be able to construct a solution which is valid at all times for low momentum fluctuations, as a function of the background fields.
The way this is done is called in the literature "the matching procedure" \cite{Hoyos:2012xc}. The idea is to match two solutions, which are valid at different regions of the space of parameters.
For low enough momentum the two regions overlap and cover together the full space of parameters. Then one can match the two solutions at the overlapping region and by that construct a solution which is valid everywhere.

The first solution is a perturbative expansion in momentum. We will show that the small expansion parameter is the momentum over the inverse comoving Hubble radius. For a given momentum, this perturbative solution will always break down at sufficiently early times when the comoving Hubble radius is large.
The second solution is valid at early times, in the quasi-de Sitter region. Using the slow-roll approximation one can solve for the fluctuation and impose the Bunch-Davies boundary conditions at early times \cite{Maldacena:2002vr}.
The perturbative solution is not valid at very early times, but for low enough momentum its region of validity extends all the way to the quasi-de Sitter regime, and the two solutions overlap.

We start by introducing fluctuations over the homogenous background \eqref{homogenousSol}
\begin{eqnarray}
  ds^2  &=& -(1+2\Phi)dt^2+2e^AB_idx^idt+e^{2A}\left[ (1-2\Psi)\delta_{ij} +  E_{ij}  +\gamma_{ij}  \right]   dx^idx^j  \\
  \phi &=& \phi(t)+\varphi(t,\vec{x})
\end{eqnarray}
where
\begin{eqnarray}
  B_i &=& \partial_iB-S_i \\
  E_{ij} &\equiv& 2\partial_i\partial_j E+2 \partial _{(i}F_{j)}
\end{eqnarray}
and
\begin{eqnarray}
  \partial^iS_i &=& \partial^{i}F_i = 0 \\
  \gamma ^i _i &=& \partial^i \gamma_{ij} = 0
\end{eqnarray}
There are five scalar perturbations $\varphi,\Phi,B,\Psi,E$. Two of them can always be set to zero using a gauge transformation (see for example \cite{Baumann:2009ds}). Out of the three variables left, there is only one dynamical degree of freedom since two of the Einstein equations are constraint equations. This dynamical fluctuation is given by the following gauge invariant variable
\begin{equation}\label{zeta}
  \zeta = \Psi + \frac{\dot{A}}{\dot{\phi}}\varphi
\end{equation}
which is called the \emph{comoving curvature perturbation}.
The vector perturbations $S_i$ and $F_i$ are not generated by inflation and in any case can be removed by an appropriate gauge transformation.
The tensor perturbation $\gamma_{ij}$ is transverse and traceless, and it is gauge invariant. We will discuss tensor perturbations in section \ref{tensor}.

We now concentrate on the scalar fluctuation and specialize to three spatial dimensions. The action for the gauge invariant scalar perturbation $\zeta$ to second order is \cite{Maldacena:2002vr}
\begin{eqnarray}\label{flucAction}
  S &=&  \int dt d^3x \frac{1}{2} \frac{\dot{\phi}^2}{\dot{A}^2} \left[ e^{3A} \dot{\zeta}^2 -e^{A} (\partial \zeta)^2 \right]
\end{eqnarray}
At this point it will be useful to introduce conformal time, which is defined by
\begin{eqnarray}
  d\tau &=& e^{-A} dt
\end{eqnarray}
Let us also define the following functions of the background fields
\begin{eqnarray}\label{ep}
  \ep &\equiv& \frac{1}{2}\frac{\dot{\phi}^2}{\dot{A}^2} = - \frac{\ddot{A}}{\dot{A}^2} =\frac{1}{2} \frac{ {\phi '} ^2  } {{A'} ^2} = 1- \frac{A''}{A'^2}\\
\label{eta}
  \eta &\equiv& \frac{\dot{\ep}}{\dot{A}\ep} = \frac{\ep '}{A' \ep }\\
\label{kappa}
  \kappa &\equiv& \frac{\dot{\eta}}{\dot{A}\eta} = \frac{\eta '}{A' \eta }
\end{eqnarray}
where $'$ denotes a derivative with respect to the conformal time.
The action of the fluctuation \eqref{flucAction} can now be written as
\begin{eqnarray}\label{flucAction2}
  S &=&  \int d\tau d^3x \; y \left[  \zeta '^2 - (\partial \zeta)^2 \right]
\end{eqnarray}
where we defined the variable\footnote{$\zeta$ and $y$ are related to the Mukhanov variable $\upsilon=z \zeta$ where $y=\frac{1}{2} z^2$.}
\begin{eqnarray}\label{yVariable}
  y &\equiv& \ep e^{2A}
\end{eqnarray}
The equation of motion of $\zeta$ in momentum space is then
\begin{eqnarray}\label{EOMzeta}
  \left( y \zeta ' \right)' +k^2 y \zeta &=& 0
\end{eqnarray}
The rest of this section is dedicated to solving this equation in different regions of the space of parameters, match the solutions and construct a solution which is valid at all times.

\subsection{Low momentum solution}

We start by solving equation \eqref{EOMzeta} perturbatively in small momentum. The solution takes the following form
\begin{eqnarray}\label{shSol}
   \zeta_{pert}(\tau) &=&  \sum_{n=0}^{\infty} \left[ D_0 U_n (\tau) +D_1 V_n(\tau)  \right] k^{2n}
\end{eqnarray}
where $U_0$ and $V_0$ are the solutions to the zero momentum equation
\begin{eqnarray}
  U_0 &=& 1 \\
  \label{v0} V_0 &=& \int d\tau \frac{1}{y}
\end{eqnarray}
and
\begin{equation}
  F_{n+1} = - \int d\tau \frac{1}{y} \int d\tau \; y \; F_n
\end{equation}
$F_n$ stands for either $U_n$ and $V_n$.

We want to determine the region of validity of the perturbative solution \eqref{shSol}.
However, since we use a dimensionful parameter, $\tau$, as the time coordinate it is not clear from \eqref{EOMzeta} what is precisely the dimensionless expansion parameter.
For that purpose we would like to find a dimensionless variable which can serve as a time coordinate and express equation \eqref{EOMzeta} in terms of this variable.
Inflationary universes only expand\footnote{There are other cosmological scenarios in which this is not true, see for example \cite{Creminelli:2007aq}.} and therefore the scale factor is a monotonically increasing function of $\tau$.
In addition, the scale factor $A(\tau)$ is dimensionless, and for this two reasons we can use it as a time coordinate.
In terms of the scale factor as the coordinate the equation of motion now takes the form
\begin{equation}\label{EOMA}
  \partial_A^2 \zeta +\frac{\partial_A(y A')}{A'} \partial_A \zeta +\left( \frac{k}{A'} \right)^2 \zeta =0
\end{equation}
$\partial_A$ is a derivative with respect to $A$. In this form everything is dimensionless and it is transparent that the dimensionless expansion parameter is
\begin{equation}\label{dep}
  \left( \frac{k}{A'} \right)^2
\end{equation}
$A'$ is the \emph{inverse comoving Hubble radius} which increase during inflation. Therefore the low momentum approximation is more accurate at late times but loses its validity at early times. More precisely, the expansion is valid in the superhorizon region where $k<A'$.

The solutions \eqref{shSol} represent a family of solutions described by the two constants $D_0$ and $D_1$. In order to determine the unique solution one also has to specify what are the boundary conditions. This solution is not valid at very early times and therefore one cannot use it directly to impose the Bunch-Davies boundary conditions. However, since there is an overlap between the quasi-de Sitter and the superhorizon regions, we can match the two solutions, and by that determine the perturbative solution uniquely.

\subsection{Quasi-de Sitter region}

In this subsection we review the systematic slow-roll expansion of the solution in the quasi-de Sitter region \cite{Baumann:2009ds}. In the slow-roll regime the parameters \eqref{ep}-\eqref{kappa} are small and one can solve for the background fields. Solving \eqref{ep} for $A$
\begin{eqnarray}
   A(\tau)&\simeq&-(1+\ep)\ln(-H \tau)
\end{eqnarray}
In the same way one can solve \eqref{eta} for $\ep$ and \eqref{kappa} for $\eta$
\begin{eqnarray}
  \ep &=& (-H \tau)^{\eta} \\
  \eta &=& (-H \tau)^{\kappa}
\end{eqnarray}
The $y$-variable therefore takes the form
\begin{eqnarray}\label{ySlowRoll}
  y &=& (-H \tau)^{1-2\nu}
\end{eqnarray}
where
\begin{eqnarray}\label{nu}
  \nu &\equiv& \frac{3}{2} +\ep +\frac{1}{2} \eta
\end{eqnarray}
In the rest of this paper we refer to $\epsilon$ and $\eta$ as the variables \eqref{ep}-\eqref{eta} evaluated in the slow-roll regime.

The equation of motion \eqref{EOMzeta} can now be solved exactly to give
\begin{eqnarray}
  \zeta_{qdS}(\tau) &=& (-H \tau )^{\nu} \left[ \gamma H^{(1)}_{\nu}(-k\tau)    +\beta H^{(2)}_{\nu}(-k\tau)  \right]
\end{eqnarray}
where $H_{\nu}^{1}$ $\left(H_{\nu}^{2}\right)$ is the Hankel function of the first (second) kind.
To impose the Bunch-Davies boundary condition at early times we consider the limit (see appendix \ref{Hankel} for the properties of the Hankel function)
\begin{eqnarray}
  \lim_{k\tau \rightarrow -\infty}\zeta_{qdS}(\tau) &=& \sqrt{\frac{2}{\pi}} (-H \tau)^{\nu-\frac{1}{2}} \left[ \gamma \frac{1}{\sqrt{k}} e^{-ik\tau}  +\beta \frac{1}{\sqrt{k}} e^{ik\tau}  \right]
\end{eqnarray}
Comparing this to the solution in de Sitter space \cite{Baumann:2009ds}
\begin{eqnarray}
  \lim_{k\tau\rightarrow -\infty} \zeta(\tau) &=& \frac{1}{2} (-H\tau)^{\nu-\frac{1}{2}} \frac{1}{\sqrt{k}}e^{-ik\tau}
\end{eqnarray}
we find
\begin{eqnarray}
  \gamma &=& \frac{1}{2}\sqrt{\frac{\pi}{2}} ,\quad \beta = 0
\end{eqnarray}
We conclude that the solution in the quasi-de Sitter region is
\begin{eqnarray}\label{qdSsol}
  \zeta_{qdS}(\tau) &=& \frac{1}{2}\sqrt{\frac{\pi}{2}}   (-H \tau )^{\nu}   H^{(1)}_{\nu}(-k\tau)
\end{eqnarray}

\subsection{The matching}

The final step in constructing the full solution at low momentum is to match the perturbative solution \eqref{shSol} with the solution in the quasi-de Sitter region \eqref{qdSsol}. This procedure fixes uniquely the coefficients $D_0$ and $D_1$ in \eqref{shSol} and therefore determine the solution at all times.

The overlapping region between the two solutions is defined to be where both the slow-roll and low-momentum approximations are valid. Taking the low momentum limit of the quasi-de Sitter solution \eqref{qdSsol} should therefore be equal to evaluating the perturbative solution \eqref{shSol} in the quasi-de Sitter region.
The low-momentum limit of the quasi-de Sitter solution \eqref{qdSsol} takes the form
\begin{eqnarray}\label{qdSLow}
  \zeta_{qdS}(\tau) &=& \sqrt{\frac{\pi}{8}}  \left(\frac{k}{H}\right)^{-\nu}  \left[  b_{\nu} + a_{\nu} (-k \tau)^{2\nu} \right]
\end{eqnarray}
$a_{\nu}$ and $b_{\nu}$ are numerical constants depending only on the rank of the Hankel function (see appendix \ref{Hankel}).

On the other hand, evaluating the perturbative solution \eqref{shSol} in the slow-roll regime is more involved. The perturbative solution is given by integrals over functions of the background fields. Since we don't know the precise form of the background solution we cannot evaluate these integrals (and even if we knew the exact homogenous solution the integrals may not necessarily be solvable). Nevertheless, it is possible to approximate the integrands in the slow-roll regime and then solve the integrals. In \cite{Hoyos:2013gma} it was argued that this procedure gives the correct result up to some mixing terms between the two series of the perturbative solution.

To be more concrete let us put an example. Evaluating $V_0$ \eqref{v0} by approximating the integrand in the slow-roll regime gives
\begin{equation}
  V_0 \sim \left(  -H\tau \right)^{2\nu}
\end{equation}
However, evaluating $V_0$ explicitly and then taking the slow-roll limit will generically give
\begin{equation}\label{v0mixing}
  V_0 \sim 1+ \left(  -H\tau \right)^{2\nu}
\end{equation}
as was argued in \cite{Hoyos:2013gma}. The constant piece in \eqref{v0mixing} then mixes with the constant term in the first series, $D_0$. Based on generic examples, it was argued in \cite{Hoyos:2013gma} that such mixing terms generically appear in the solution. For more details and discussions about the mixing issue we refer the reader to \cite{Hoyos:2013gma}.

The perturbative solution evaluated in the quasi-de Sitter region and including possible mixing terms is then given by
\begin{eqnarray}\label{pertSol}
  \zeta_{pert}(\tau) &=& \left[D_0+c_2(1+\dots)D_1\right] - \frac{D_1+c_1(k^2+\dots)D_0}{2\nu H} (-H\tau)^{2\nu}
\end{eqnarray}
$c_1$ and $c_2$ stand for the mixing terms between the two series.
Matching the two solutions \eqref{qdSLow} and \eqref{pertSol} we find
\begin{eqnarray}
  D_0 &=& \sqrt{\frac{\pi}{8}} b_{\nu} \left(\frac{k}{H}\right)^{-\nu}
  \frac{
  1-c_2(1+\dots)P(k)
  }
  {
  1-c_1 c_2 (k^2+\dots)
  }\\
\label{bc}
  \frac{D_1}{D_0} &=& \frac{
  P(k)-c_1(k^2+\dots)
  }
  {
  1-c_2P(k)(1+\dots)
  }
\end{eqnarray}
where
\begin{equation}\label{P}
  P(k) \equiv -2\nu \frac{a_{\nu}}{b_{\nu}} H  \left(\frac{k}{H}\right)^{2\nu}
\end{equation}
This procedure uniquely determines the perturbative solution.

\section{The dual flow}

The dS/CFT correspondence postulates that the wavefunction of the universe which is asymptotically de Sitter space is equal to the partition function of a dual field theory. The precise dictionary is \cite{Maldacena:2002vr}
\begin{eqnarray}\label{dictionary2}
  \Psi[g] &=& Z[g]
\end{eqnarray}
where the left hand side is the wavefunction of the universe for a given three metric $g$ and the right side is the partition function of some dual three dimensional field theory. The wavefunction of the universe is given by
\begin{eqnarray}
  \Psi[g] &=& \int \mathcal{D}[G] e^{i S[G]}
\end{eqnarray}
where $G$ is a four metric.
We will use the classical gravity approximation \cite{Maldacena:2002vr}
\begin{eqnarray}\label{clWF}
  \Psi &\simeq& e^{i S_{cl}}
\end{eqnarray}
$g$ is the three metric of the spatial slice at asymptotically late times $\tau \rightarrow 0$. In order to study the dual field theory using the dictionary \eqref{dictionary2} we therefore first need to discuss the asymptotic form of the solution for the fluctuation $\zeta$.

\subsection{Asymptotically late times}

At asymptotically late times the universe is a de Sitter space with a Hubble constant $H_0$.
In this region the solution is very well-approximated by the perturbative expansion \eqref{shSol}.
In analogy to the quasi-de Sitter region \eqref{ySlowRoll}-\eqref{nu}, the $y$-variable at the late $dS$ is given by
\begin{equation}\label{ydS}
  y= \left( -H_0 \tau \right)^{1-2\nu_0}
\end{equation}
with
\begin{equation}\label{nuzero}
  \nu_0 \equiv \frac{3}{2} +\epsilon_0+\frac{1}{2}\eta_0
\end{equation}
$\epsilon_0$ and $\eta_0$ are the variables \eqref{ep} and \eqref{eta}, respectively, evaluated in the late de Sitter space.
They are the analogs of the primordial de Sitter's slow-roll parameters, for the late de Sitter.
To be more concrete, let us evaluate them explicitly using the asymptotic solution of the background fields \eqref{scalarSol} and \eqref{scaleFactor}. To leading order
\begin{equation}
  -H_0 \tau \simeq e^{-H_0 t}
\end{equation}
and
\begin{eqnarray}
  \epsilon_0 &\simeq& \frac{C^2 \Delta_{\pm}^2}{2} \left( -H_0 \tau \right)^{2\Delta_{\pm}} \\
  \eta_0 &\simeq& -2\Delta_{\pm}
\end{eqnarray}
Both roots $\Delta_{\pm}$ are positive and therefore $\ep_0$ goes to zero in the far future $\tau\rightarrow 0$. However, $\eta_0$ is finite and not necessarily small (as opposed to the slow-roll's $\eta$).
We therefore find that
\begin{equation}\label{nu0n}
  \nu_{0\pm} = \frac{3}{2}-\Delta_{\pm}=\mp \sqrt{\left(\frac{3}{2}\right)^2-\left(\frac{m}{H_0}\right)^2}
\end{equation}
Evaluating the solution for $\zeta$ \eqref{shSol} in this region \eqref{ydS} we find
\begin{eqnarray}\label{dsSol}
  \zeta _{dS} &=& D_0 - \frac{D_1}{2\nu_0H_0} (-H_0 \tau)^{2\nu_0}+\dots
\end{eqnarray}
where the dots refer to subleading corrections to both series.

Knowing the asymptotic form of the solution we are now able to calculate the gravitational wavefunction since that in the classical approximation it is given by the on-shell action \eqref{clWF}, which is a boundary term.
Using the equation of motion, the action for the fluctuation \eqref{flucAction2} can be written as a boundary term
\begin{eqnarray}\label{boundaryAction}
  S &=& \int \frac{d^3 k}{(2\pi)^3} \left[ y\zeta \zeta ' \right]_{\tau = \delta}
\end{eqnarray}
$\delta$ is a cutoff which should be taken to zero eventually.
Evaluating the action \eqref{boundaryAction} on the asymptotic solution \eqref{dsSol} we find
\begin{eqnarray}\label{action}
  \nonumber S &=& \int \frac{d^3 k}{(2\pi)^3}
  \left[ D_0D_1\left(1+\dots \right)\right. \\ &&
  \nonumber \left.
  - D_0^2 \left(-H_0\delta\right)^{-\nu_0}  \frac{H_0\left(k \delta\right)^2}{\nu_0-2} \left(1+\dots\right)\right. \\ &&
  \left.
  -D_1^2 (-H_0\delta)^{\nu_0} \frac{1}{H_0\nu_0}\left(1+\dots\right)
   \right]
\end{eqnarray}
where dots indicate higher order terms in $k\delta$.
Only the term independent of $\delta$ in the expression above \eqref{action}, which is proportional to $D_0D_1$, gives rise to the two-point function.
All the other terms are either divergent or vanish as $\delta \rightarrow 0$. The divergent terms are viewed as divergences in the field theory which should be subtracted by local counterterms \cite{Maldacena:2002vr}.
We conclude that the finite non-local part of the boundary action is
\begin{eqnarray}\label{actionNonLocal}
  S_{finite, non-local} &=& \int \frac{d^3 k}{(2\pi)^3}D_0D_1
\end{eqnarray}

\subsection{The two-point function}
Using the dictionary \eqref{dictionary2} one can calculate correlation functions in the dual theory from the gravity wavefunction. In particular, we will be interested in the two-point function of the dual scalar operator of the inflaton
\begin{eqnarray}
  \left< \mathcal{O}_{\vec{k}}\mathcal{O}_{\vec{k}'} \right> &=& \left[ \frac{\delta^2 Z}{\delta J(\vec{k}) \delta J(\vec{k}')} \right]_{J=0}
\end{eqnarray}
Using the result for the on-shell action \eqref{actionNonLocal}, the two-point function is given by
\begin{equation}\label{result2}
  \left< \mathcal{O}_{\vec{k}}\mathcal{O}_{\vec{k}'} \right> = i(2\pi)^3 \delta\left(\vec{k}+\vec{k}'\right) \left[ \frac{\delta^2 \left(D_0D_1\right)}{\delta J(\vec{k}) \delta J(\vec{k}')} \right]_{J=0}
\end{equation}

$J(\vec{k})$ is the source for the dual operator and it is defined using the asymptotic form of the bulk scalar field \eqref{dsSol}.
More precisely, the source $J(\vec{k})$ is defined to be the coefficient of the leading term in the expansion \eqref{dsSol}. Which of the two terms in \eqref{dsSol} is leading depends on the sign of $\nu_0$. For the lower root $\nu_{0-}>0$ the second solution in \eqref{dsSol} goes to zero at late times ($\tau\rightarrow 0$) and therefore $D_0$ is identified as the source for the dual scalar operator
\begin{equation}\label{Jm}
  J_{-}(\vec{k})=D_0(\vec{k})
\end{equation}
For the upper root $\nu_{0+}<0$ the second solution in \eqref{dsSol} diverges at late times and therefore $D_1$ is proportional to the source
\begin{equation}\label{Jp}
  J_{+}(\vec{k})=- \frac{1}{2\nu_0H_0}D_1(\vec{k})
\end{equation}
Later on we will interpret these two cases as two phases of the dual field theory: the case $\nu_{0-}$ will correspond to explicit breaking of conformal symmetry, while the case $\nu_{0+}$ will correspond to spontaneous breaking of conformal symmetry.

Using the definitions of the source in the two cases, \eqref{Jm} and \eqref{Jp}, and the boundary condition we found \eqref{bc}, we are now able to explicitly evaluate the two-point function \eqref{result2}.

\subsection{Explicit breaking}
We begin with the case $\nu_{0-}$, for which the source $J(\vec{k})$ is equal to $D_0(\vec{k})$ and equation \eqref{result2} reduces to
\begin{equation}
  \left< \mathcal{O}_{\vec{k}}\mathcal{O}_{\vec{k}'} \right>_{-} = i(2\pi)^3 \delta\left(\vec{k}+\vec{k}'\right)
  \left( \frac{D_1}{D_0} \right)
\end{equation}
Using the boundary condition \eqref{bc} we then find
\begin{equation}\label{tpf}
  \left< \mathcal{O}_{\vec{k}}\mathcal{O}_{\vec{k}'} \right>_{-} = i(2\pi)^3 \delta\left(\vec{k}+\vec{k}'\right)
  \frac{
  P(k)-c_1(k^2+\dots)
  }
  {
  1-c_2P(k)(1+\dots)
  }
\end{equation}
The function $P(k)$ \eqref{P} is a non-integer positive power of the momentum $k$ and therefore when expanding the last expressions \eqref{tpf} in small momentum we find that the leading non-analytic piece is given by
\begin{equation}
  \left< \mathcal{O}_{\vec{k}}\mathcal{O}_{\vec{k}'} \right>_{-} =
  i(2\pi)^3 \delta\left(\vec{k}+\vec{k}'\right) P\left(k\right)+\dots
\end{equation}
The dots refer to subleading terms in momentum and analytic contributions which are viewed as contact terms in the dual field theory. The leading non-analytic contribution to the two point function is therefore
\begin{eqnarray}\label{explicitCorr}
  \left< \mathcal{O}_{\vec{k}}\mathcal{O}_{\vec{k}'} \right>_{-} &=&
  -i(2\pi)^3 \delta\left(\vec{k}+\vec{k}'\right)  \frac{2\nu a_{\nu}}{b_{\nu}} H  \left(\frac{k}{H}\right)^{2\nu}
\end{eqnarray}
The two-point function is of the form
\begin{eqnarray}
  \left< \mathcal{O}\mathcal{O} \right>_{-} &\sim& k^{3-2\lambda}
\end{eqnarray}
where
\begin{eqnarray}
  \lambda &=& -\epsilon -\frac{1}{2} \eta
\end{eqnarray}
This is a correlator of a nearly marginal operator of dimension
\begin{eqnarray}
  \Delta &=& 3-\lambda
\end{eqnarray}
in a three dimensional field theory.

\subsection{Spontaneous breaking}
Let us go on to the second case $\nu_{0+}$, in which the source $J(\vec{k})$ is proportional to $D_1(\vec{k})$. Now equation \eqref{result2} gives
\begin{equation}
  \left< \mathcal{O}_{\vec{k}}\mathcal{O}_{\vec{k}'} \right>_{+} = i(2\pi)^3 \delta\left(\vec{k}+\vec{k}'\right)
  \left( 2\nu_0 H_0 \right)^2\left( \frac{D_1}{D_0} \right) ^{- 1}
\end{equation}
Plugging in the boundary condition \eqref{bc} we find
\begin{equation}\label{correlator}
  \left< \mathcal{O}_{\vec{k}}\mathcal{O}_{\vec{k}'} \right>_{+} = i(2\pi)^3 \delta\left(\vec{k}+\vec{k}'\right)
 \left( 2\nu_0 H_0 \right)^2  \frac{ 1-c_2P(k)(1+\dots)  }{  P(k)-c_1(k^2+\dots)}
\end{equation}
Now since $P(k)\sim k^{2\nu}$, and $2\nu >2$ (see \eqref{nu}), the leading non-analytic piece in the correlator \eqref{correlator} is
\begin{equation}
  \left< \mathcal{O}_{\vec{k}}\mathcal{O}_{\vec{k}'} \right>_{+} =
 - i(2\pi)^3 \delta\left(\vec{k}+\vec{k}'\right) \frac{\left( 2\nu_0 H_0 \right)^2}{c_1}\frac{1}{k^2}+\dots
\end{equation}
where the dots again indicate subleading and analytic terms in momentum. We interpret the massless pole appearing in the two point function as the propagator of the dilaton, the Goldstone boson of the spontaneously broken conformal symmetry \cite{Hoyos:2013gma}.

\section{Gravitational waves}\label{tensor}

We now turn to study tensor fluctuations over the homogenous background. The quadratic action for the tensor fluctuation $\gamma_{ij}$ is given by \cite{Maldacena:2002vr}
\begin{eqnarray}\label{gwAction}
  S &=& \frac{1}{8} \int dt d^3 x  \left[ e^{3A} \dot{\gamma}_{ij}\dot{\gamma}_{ij} - e^{A} \partial_{\ell}\gamma_{ij}\partial_{\ell}\gamma_{ij} \right]
\end{eqnarray}
$\gamma_{ij}$ can be expanded in plane waves with definite polarization tensors
\begin{eqnarray}
  \gamma_{ij} &=& \int \frac{d^3 k}{(2\pi)^3} \sum_{s=\pm} \ep_{ij}^s(k)\gamma^s_{\vec{k}}(t) e^{i\vec{k}\vec{x}}
\end{eqnarray}
where $\ep_{ii}=k^{i}\ep_{ij}=0$ and $\ep_{ij}^s(k)\ep_{ij}^{s'}(k)=2\delta_{ss'}$. Plugging this decomposition into the action \eqref{gwAction} and translating to conformal time results in
\begin{eqnarray}
  S &=& \frac{1}{4} \int d\tau  \frac{d^3 k}{(2\pi)^3}  e^{2A} \sum_{s=\pm}\left[  \left({\gamma^{s}_{\vec{k}}}'\right)^2 -k^2 \left({\gamma^{s}_{\vec{k}}}\right)^2  \right]
\end{eqnarray}
The equation of motion for each polarization mode is then
\begin{eqnarray}
  \left( e^{2A}{\gamma^{s}_{\vec{k}}}' \right)' +k^2 e^{2A}\gamma^{s}_{\vec{k}}   &=& 0
\end{eqnarray}
This equation is very similar to the equation of motion of the scalar mode \eqref{EOMzeta} with the $y$ variable \eqref{yVariable} replaced by $y_{t}\equiv e^{2A}$. Therefore we can solve in the same way for the tensor modes and find the correlator of the dual operator (the energy-momentum tensor).

In the quasi-de Sitter region $y_{t}$ takes the form
\begin{eqnarray}
  y_{t} &=&  (-H \tau)^{1-2\sigma}
\end{eqnarray}
with
\begin{eqnarray}
  \sigma &\equiv& \frac{3}{2} +\ep
\end{eqnarray}
From here on everything goes through precisely in the same way as for the scalar mode with the replacement $\nu \rightarrow \sigma $.
However, in that case we have
\begin{equation}
  \sigma_0 \simeq \frac{3}{2}
\end{equation}
in the late $dS$ space. $\sigma_0$ is positive for both cases $\Delta_{\pm}$ and therefore the two-point function of the dual energy-momentum tensor is the same in both explicit and spontaneous breaking phases, and it takes the form
\begin{equation}
  \left< T_{\vec{k}}^{s} T_{\vec{k'}}^{s'}  \right> = -i(2\pi)^3 \delta^{ss'} \delta\left(\vec{k}+\vec{k}'\right) \frac{2 \sigma a_{\sigma}}{b_{\sigma}} H \left( \frac{k}{H} \right) ^{2\sigma}
\end{equation}

The momentum dependence of the two-point function is
\begin{eqnarray}
  \left< T_{\vec{k}}^{s} T_{\vec{k'}}^{s'} \right> &\sim& k^{3-2\lambda_t}
\end{eqnarray}
where
\begin{eqnarray}
  \lambda_t &=& -\epsilon
\end{eqnarray}
This is a correlator of a nearly marginal operator of dimension
\begin{eqnarray}
  \Delta &=& 3-\lambda_t
\end{eqnarray}
in a three dimensional field theory.

\section{Concluding remarks and future directions}

In the present paper we have studied the dual flow to single-field slow-roll inflation. We found the solution for the inflationary fluctuation at all times using a low momentum approximation and a matching procedure between a perturbative and a slow-roll expansions.
Using this solution we were able to apply the postulate of the dS/CFT correspondence, in the same way as in AdS, and calculate the two-point function of the dual operator to the inflaton.
We found that the dual flow admits two phases of explicit and spontaneous breaking of conformal symmetry (or of the de Sitter isometries).
The two-point function of the dual operator in the explicit breaking phase is
\begin{eqnarray}
  \left< \mathcal{O}\mathcal{O} \right>_{explicit} &\sim& k^{3-2\lambda}
\end{eqnarray}
$\lambda$ is the anomalous dimension of the dual irrelevant operator and is related to the slow-roll parameters by
\begin{eqnarray}
  \lambda &=& -\epsilon -\frac{1}{2} \eta
\end{eqnarray}
Since the slow-roll parameters are small, the dual operator is nearly marginal (or slightly irrelevant).
The two-point function of the dual operator in the spontaneous breaking phase, on the other hand, is a massless pole
\begin{eqnarray}
  \left< \mathcal{O}\mathcal{O} \right>_{spontaneous} &\sim& \frac{1}{k^2}
\end{eqnarray}
which is associated with the dilaton - the Goldstone boson of spontaneously broken conformal symmetry.

From the bulk point of view, the two phases are related to the behavior of the homogenous background inflaton, which has two solutions.
The solution which goes faster to zero at late times corresponds to the spontaneous breaking phase while the other solution corresponds to the explicit breaking phase.
It is therefore necessary to fine-tune the leading background solution at late times to zero in order to land in the spontaneous breaking phase.
Since the power spectrum of fluctuations in the bulk is schematically given by
\begin{equation}
  \mathcal{P} \sim \frac{1}{\left< \mathcal{O}\mathcal{O} \right>}
\end{equation}
we conclude that in a generic, non-fine-tuned, case slow-roll inflation generates a nearly scale invariant (or quasi-marginal) power spectrum
\begin{equation}
  \mathcal{P} \sim \frac{1}{k^{3-2\lambda}}
\end{equation}
However, we find a possible fine-tuned situation in which
\begin{equation}
  \mathcal{P} \sim k^2
\end{equation}
We conclude that slow-roll inflation does not guarantees a nearly scale invariant spectrum, but the spectrum is generically nearly scale invariant except in fine-tuned situations.

The appearance of the spontaneous breaking phase is a direct consequence of the mixing between the two fluctuating modes in the bulk. The matching procedure mix between the normalizable and the non-normalizable modes such that each of the modes at late times is a superposition of the two modes in the quasi-de Sitter region. The mixing effect induce a dramatic change in the spontaneous breaking phase, but does not affect the power spectrum in the explicit breaking phase (at least to leading order in momentum). Together with the fact that the spontaneous breaking phase is fine-tuned this is consistent with the experimental measurement of a nearly scale invariant power spectrum. However, it opens a window for a variety of other effects of the mixing on other cosmological observables. For example, it may be very interesting to study the effect of the mixing between the two modes in the bulk on the bispectrum (the three point function) and see if it has a unique signature on the shape of non-gaussianities.

Further proposals for future directions are in place. The effective field theory of inflation was first introduced in \cite{Cheung:2007st} using the St\"{u}ckelberg trick. The Goldstone boson is introduced by performing a broken time diffeomorphism and its effective action parameterizes the most general theory of fluctuations around the quasi-de Sitter background.
The effective action was then used to put constraints on inflation, find consistency conditions and discover new cosmological models.
It would be interesting to relate the Goldstone boson in the bulk to the spurion field of the boundary conformal symmetry (the dilaton in the case of spontaneous breaking). In particular, the constraints on the dilaton effective action that were found in the seminal works \cite{Komargodski:2011vj,Komargodski:2011xv} could then be translated into constrains on the effective action in the bulk.
In this context let us mention, for example, two interesting works \cite{Creminelli:2011mw,Creminelli:2012ed} which discuss the relation between conformal symmetry and inflation and which can be used for the purposes discussed above.
Another direction we would like to pursue is to borrow the effective field theory approach of inflation to holography. In the same way in which the Goldstone boson of inflation was introduced, one can use a broken radial diffeomorphism along the holographic direction to introduce a Goldstone boson for holography. The effective action of this Goldstone boson will then parameterize the most general asymptotically AdS holographic background. Such effective action can be useful for discovering new dualities and to better understand the holographic correspondence between gravity and quantum field theory.
Finally, multi-trace deformations were intensively studied in AdS/CFT. In a similar way, it would be interesting to study multi-trace deformations in dS/CFT and their effect on inflation (see \cite{Das:2013qea} for a recent work on double-trace deformations in dS/CFT).

\acknowledgments
I would like to thank Carlos Hoyos, Cobi Sonnenschein, Shimon Yankielowicz and Sunny Itzhaki for many useful discussions.
U.K. is supported by the Lev Zion fellowship of the Council for Higher Education (Israel).
This work was supported in part by the Israel Science Foundation (grant number 1468/06).

\appendix

\section{The Hankel function}\label{Hankel}
Hereby we give some properties of the Hankel function of the two kinds. The limit at infinty
\begin{eqnarray}
  \lim_{z\rightarrow \infty} H_{\theta}^{(1,2)}(z) &=& \sqrt{\frac{2}{\pi}}\frac{1}{\sqrt{z}}e^{\pm i z}
\end{eqnarray}
The limit at zero
\begin{eqnarray}
  \lim_{z\rightarrow 0} H_{\theta}^{(1)}(z) &=&   + b_{\theta} z^{-\theta}  +        \qquad    a_{\theta} z ^{\theta}    \\
  \lim_{z\rightarrow 0} H_{\theta}^{(2)}(z) &=&   - b_{\theta} z^{-\theta}  - e^{2\pi\theta i} a_{\theta} z ^{\theta}
\end{eqnarray}
where
\begin{eqnarray}
  b_{\theta} &\equiv& -\frac{i 2^{\theta} \Gamma(\theta) }{\pi} \\
  a_{\theta} &\equiv& \frac{2^{-\theta} \left[ 1+i \cot(\pi\theta) \right]}{\Gamma(1+\theta)}
\end{eqnarray}


\begin{thebibliography}{}





\bibitem{Witten:2001kn}
  E.~Witten,
  ``Quantum gravity in de Sitter space,''
  hep-th/0106109.




\bibitem{Strominger:2001pn}
  A.~Strominger,
  ``The dS / CFT correspondence,''
  JHEP {\bf 0110}, 034 (2001)
  [hep-th/0106113].




\bibitem{Maldacena:2002vr}
  J.~M.~Maldacena,
  JHEP {\bf 0305}, 013 (2003)
  [astro-ph/0210603].




\bibitem{Strominger:2001gp}
  A.~Strominger,
  ``Inflation and the dS / CFT correspondence,''
  JHEP {\bf 0111}, 049 (2001)
  [hep-th/0110087].











\bibitem{Larsen:2002et}
  F.~Larsen, J.~P.~van der Schaar and R.~G.~Leigh,
  ``De Sitter holography and the cosmic microwave background,''  JHEP {\bf 0204}, 047 (2002)  [hep-th/0202127].  


\bibitem{Larsen:2003pf}
  F.~Larsen and R.~McNees,
  ``Inflation and de Sitter holography,''  JHEP {\bf 0307}, 051 (2003)  [hep-th/0307026].  


\bibitem{vanderSchaar:2003sz}
  J.~P.~van der Schaar,
  ``Inflationary perturbations from deformed CFT,''  JHEP {\bf 0401}, 070 (2004)  [hep-th/0307271].  

\bibitem{Larsen:2004kf}
  F.~Larsen and R.~McNees,
  ``Holography, diffeomorphisms, and scaling violations in the CMB,''
  JHEP {\bf 0407}, 062 (2004)
  [hep-th/0402050].







\bibitem{Schalm:2012pi}
  K.~Schalm, G.~Shiu and T.~van der Aalst,
  ``Consistency condition for inflation from (broken) conformal symmetry,''
  arXiv:1211.2157 [hep-th].


\bibitem{Bzowski:2012ih}
  A.~Bzowski, P.~McFadden and K.~Skenderis,
  ``Holography for inflation using conformal perturbation theory,''
  arXiv:1211.4550 [hep-th].



\bibitem{McFadden:2013ria}
  P.~McFadden,
  ``On the power spectrum of inflationary cosmologies dual to a deformed CFT,''
  arXiv:1308.0331 [hep-th].




\bibitem{Kiritsis:2013gia}
  E.~Kiritsis,
  ``Asymptotic freedom, asymptotic flatness and cosmology,''
  arXiv:1307.5873 [hep-th].




\bibitem{Nastase:2011qz}
  H.~Nastase and J.~Sonnenschein,
  ``Towards brane-antibrane inflation in type $II_A$: The Holographic MQCD model,''
  JHEP {\bf 1202}, 040 (2012)
  [arXiv:1109.6813 [hep-th]].





\bibitem{Sinha:2010pm}
  A.~Sinha,
  ``On higher derivative gravity, $c$-theorems and cosmology,''
  Class.\ Quant.\ Grav.\  {\bf 28}, 085002 (2011)
  [arXiv:1008.4315 [hep-th]].









\bibitem{Maldacena:1997re}
  J.~M.~Maldacena,
  ``The Large N limit of superconformal field theories and supergravity,''
  Adv.\ Theor.\ Math.\ Phys.\  {\bf 2}, 231 (1998)
  [hep-th/9711200].

\bibitem{Aharony:1999ti}
  O.~Aharony, S.~S.~Gubser, J.~M.~Maldacena, H.~Ooguri and Y.~Oz,
  ``Large N field theories, string theory and gravity,''
  Phys.\ Rept.\  {\bf 323}, 183 (2000)
  [hep-th/9905111].






\bibitem{Klebanov:1999tb}
  I.~R.~Klebanov and E.~Witten,
  ``AdS / CFT correspondence and symmetry breaking,''
  Nucl.\ Phys.\ B {\bf 556}, 89 (1999)
  [hep-th/9905104].

\bibitem{Freedman:1999gp}
  D.~Z.~Freedman, S.~S.~Gubser, K.~Pilch and N.~P.~Warner,
  ``Renormalization group flows from holography supersymmetry and a c theorem,''
  Adv.\ Theor.\ Math.\ Phys.\  {\bf 3}, 363 (1999)
  [hep-th/9904017].








\bibitem{Maldacena:2011nz}
  J.~M.~Maldacena and G.~L.~Pimentel,
  ``On graviton non-Gaussianities during inflation,''
  JHEP {\bf 1109}, 045 (2011)
  [arXiv:1104.2846 [hep-th]].



\bibitem{Mata:2012bx}
  I.~Mata, S.~Raju and S.~Trivedi,
  ``CMB from CFT,''
  arXiv:1211.5482 [hep-th].









\bibitem{Hoyos:2012xc}
  C.~Hoyos, U.~Kol, J.~Sonnenschein and S.~Yankielowicz,
  ``The a-theorem and conformal symmetry breaking in holographic RG flows,''
  arXiv:1207.0006 [hep-th].




\bibitem{Hoyos:2013gma}
  C.~Hoyos, U.~Kol, J.~Sonnenschein and S.~Yankielowicz,
  ``The holographic dilaton,''
  arXiv:1307.2572 [hep-th].




\bibitem{Bajc:2013wha}
  B.~Bajc and A.~R.~Lugo,
  ``On the matching method and the Goldstone theorem in holography,''
  arXiv:1304.3051 [hep-th].










\bibitem{Baumann:2009ds}
  D.~Baumann,
  ``TASI Lectures on Inflation,''
  arXiv:0907.5424 [hep-th].








\bibitem{Creminelli:2007aq}
  P.~Creminelli and L.~Senatore,
  ``A Smooth bouncing cosmology with scale invariant spectrum,''
  JCAP {\bf 0711}, 010 (2007)
  [hep-th/0702165].














\bibitem{Cheung:2007st}
  C.~Cheung, P.~Creminelli, A.~L.~Fitzpatrick, J.~Kaplan and L.~Senatore,
  ``The Effective Field Theory of Inflation,''
  JHEP {\bf 0803}, 014 (2008)
  [arXiv:0709.0293 [hep-th]].






\bibitem{Komargodski:2011vj}
  Z.~Komargodski and A.~Schwimmer,
  ``On Renormalization Group Flows in Four Dimensions,''
  JHEP {\bf 1112}, 099 (2011)
  [arXiv:1107.3987 [hep-th]].

\bibitem{Komargodski:2011xv}
  Z.~Komargodski,
  ``The Constraints of Conformal Symmetry on RG Flows,''
  JHEP {\bf 1207}, 069 (2012)
  [arXiv:1112.4538 [hep-th]].












\bibitem{Creminelli:2011mw}
  P.~Creminelli,
  ``Conformal invariance of scalar perturbations in inflation,''
  Phys.\ Rev.\ D {\bf 85}, 041302 (2012)
  [arXiv:1108.0874 [hep-th]].







\bibitem{Creminelli:2012ed}
  P.~Creminelli, J.~Norena and M.~Simonovic,
  ``Conformal consistency relations for single-field inflation,''
  JCAP {\bf 1207}, 052 (2012)
  [arXiv:1203.4595 [hep-th]].









\bibitem{Das:2013qea}
  D.~Das, S.~R.~Das and G.~Mandal,
  ``Double Trace Flows and Holographic RG in dS/CFT correspondence,''
  arXiv:1306.0336 [hep-th].














\end{thebibliography}
\end{document}